# Dielectric anomalies and magnetodielectric coupling behavior of single crystalline $Ca_3Co_2O_6$, a geometrically frustrated magnetic spin-chain system

Tathamay Basu,[#] Kartik K. Iyer, P. L. Paulose and E. V. Sampathkumaran

*Tata Institute of Fundamental Research, Homi Bhabha Road, Colaba, Mumbai 400005, India*

**Abstract**

The dielectric behaviour of the single crystals of the spin-chain system $Ca_3Co_2O_6$, undergoing geometrically frustrated antiferromagnetic ordering below 25 K, has been investigated as a function of temperature and magnetic field (*H*) and compared with magnetization (*M*) behaviour. The results provide evidence for anisotropic magnetodielectric (MDE) coupling in this compound. Ac susceptibility exhibits a strong frequency dependence for *H//c* with changes of this feature with the application of external *dc* magnetic field. No feature in *ac* susceptibility could be observed for $H \perp c$, thereby providing evidence for strong anisotropic spin-glass behaviour. Interestingly, the strong frequency dependence in dielectric is present for both the crystallographic directions (*E//c* and $E \perp c$ where *E* is the electric field) with a negligible influence of *H*, despite the existence of MDE coupling. This result appears to suggest different dynamics of electric dipole and spin-glass, although they are coupled with each other. In addition, interestingly, there is also a step at *one-third* of high-(magnetic) field value of MDE at some temperatures tracking a similar step in *M(H)*. This work also confirms following unusual features reported on polycrystals, when measured along *c*-axis. The dielectric constant exhibits a broad peak around 50-130 K when the electric field is applied along the spin-chain (crystallographic '*c*' direction); however this feature is essentially absent for the perpendicular orientation. This finding supports the role of incipient spin-chain ordering to induce magnetodielectric coupling. There is a signature of 'magnetoelectric phase coexistence' when the magnetic field is applied along the spin-chain irrespective of the direction of applied electric field.

**Key words:** $Ca_3Co_2O_6$, dielectric, magnetodielectric coupling, geometrical frustration, spin-chain

[#] Present address: Experimental Physics V, Center for Electronic Correlations and Magnetism, University of Augsburg, Augsburg 86159, Germany

# 1. Introduction

Among the spin-chain compounds crystallizing in the $K_4CdCl_6$-type rhombohedral structure (space group $R\bar{3}c$), $Ca_3Co_2O_6$ [1, 2] has drawn considerable interest [see, for instance, Refs. 3-14] in the field of geometrically frustrated magnetism. This compound consists of infinite chains (along '$c$') of alternating face-sharing $CoO_6$ distorted octahedral and $CoO_6$ distorted trigonal prisms (TP), where Co in octahedral site is in low-spin state due to crystal field effect and is non-magnetic (S=0) and Co in TP site is magnetic with high-spin state (S=1) [1, 11]. The chains are separated by Ca atom in the '$ab$' plane. This compound undergoes a complex magnetic ordering [4], with transitions at ($T_N=T_1=$) 24 K, ~10 K ($T_2$) with a huge frequency dependent ac susceptibility ($\chi$) behavior, and near 7 K ($T_3$) [5-7]. This compound also attracted some interest [12, 13] even in the field of magnetodielectric (MDE) coupling, but such studies were restricted to temperature region in the vicinity of $T_N$, bringing out magnetoelectric coupling. Recently, we have reported [14] that this compound exhibits dielectric anomalies, not only below $T_N$, but also well above $T_N$, thereby bringing out an unusual role played by spin-chains. Such a report was based on studies on polycrystals and it is desirable to carry out such investigations on single crystals over a wide temperature ($T$) range. Besides, the dielectric behaviour reported in the close vicinity of magnetic transition differs qualitatively among the previous reports [12-14]. In addition, we reported [14] evidence for a fascinating phenomenon "magnetodielectric-phase-co-existence" for the polycrystalline form, that is, a mixture of high-field (magnetic) phase and virgin phase after a field cycling across magnetic-field-induced magnetic transition. We therefore consider it important to perform careful dielectric studies on single crystals. This is the primary motivation with which we have taken up $ac$ and $dc$ magnetization ($M$) and dielectric studies in the presence of magnetic field ($H$) over a wide temperature range on single crystals.

## 2. Experimental Details:

Single crystals of $Ca_3Co_2O_6$ compound were grown by a flux growth method, as described earlier [9]. High purity (>>99.9%) $CaCO_3$ and $Co_3O_4$ in proper ratio were mixed thoroughly. This mixture and the flux ($K_2CO_3$) were mixed in the ratio of 1:20. This mixture were heated to 920 C at a rate of 300 C/h and then held for 24 h at this temperature. The specimen was then cooled down at a rate of 100 C/h to 570 C and finally to room temperature with a rate of 300 C/h. The heat-treated mixture was placed in distilled water to remove $K_2CO_3$ and the needle-shaped crystals were extracted carefully from this solution. The sintering temperature and the cooling rate were optimized from various trials to get the longest (about 3mm) possible needles of crystals. Some single crystals were crushed into powder and characterized by x-ray diffraction (Cu-K$_\alpha$). The energy dispersive x-ray analysis has been carried out to confirm the stoichiometry of the desired compound using a scanning electron microscope and the backscattered electron images have been obtained to check the homogeneity and stoichiometry of the crystal. The *dc* and *ac* magnetization studies were measured with the help of a Superconducting Quantum Interference Device (Quantum Design, QD) and isothermal *M* was carried out using Vibrating Sample Magnetometer (QD). The complex dielectric measurements were performed using a LCR meter (Agilent E4980A) with a homemade sample holder integrated to a commercial Physical Properties Measurements System (PPMS). In all the measurements, the investigations were done after cooling the sample from 200 K in the absence of any external magnetic field ("ZFC" mode), unless specifically stated otherwise elsewhere in the article. The rate of change of magnetic field was maintained at 70 Oe/s in all the cases.

3. Results and discussion

*Dc magnetization*

Though the *dc* magnetization of single crystals has been already reported by us in the past [9], agreeing with the literatures [7, 15], we present here again for the benefit of the reader to understand subsequent discussions. Besides, the present measurements were performed on one crystal, rather than on a bunch of crystals as in our previous study [9]. As reported in the literature, there is a sharp rise near $T_1$ (= 24 K) in *M(T)* for the orientation of *H//c* followed by a peak near $T_2$ ( around 10 K), as shown in figure 1a. There is an additional shoulder near 7 K (inset of figure 1a), indicating additional magnetic anomaly at low temperatures as seen in polycrystals. The observation of equally spaced steps (~12 kOe) in the virgin curve of *M(H)* for the orientation *H//c* at 1.8 K and the complete saturation at high magnetic fields (see figure 1c) are in conformity with the literature data on the single crystal of this compound [7]. These features are absent for the other orientation *H$\perp$c* and the absolute values of *M* are much smaller, consistent with strong anisotropic nature of magnetism of this compound (see figure 1c and d). In addition, there is a broad maxima in the *M(T)* plot for orientation *H$\perp$c*, as a signature of short-range magnetic correlation due to spin-chains [15]. The almost linear *M(H)* behavior for orientation *H$\perp$c* is consistent with interchain antiferromagnetic interaction; the sharp rise below 20 K in *M(T)* plot or weak curvature/hysteresis in *M(H)* could be due to a small misalignment of the crystal.

*Ac susceptibility behavior and the influence of magnetic field*

We show *ac* χ (measured with *ac* field of 1 Oe) behavior in figure 2 in the absence and in the presence of external magnetic field for the orientation *H//c*. Such a study on the polycrystalline samples has been reported in the past literature, [6, 7]. The real (χ′) and imaginary (χ″) parts of *ac* susceptibility show a huge frequency (ν) dependent behavior in the absence of external field, as known in the past literature. Application of a magnetic field of 10 kOe suppresses the frequency dependent behavior of χ′ as well as its intensity and the peaks in *χ″(T)* are completely suppressed, as shown in figure 2b. Application of a magnetic field of 30 kOe interestingly restores the peaks in both real and imaginary parts, but at a higher temperature of 20 K, but without any apparent frequency dependence as shown in figure 2c. Further application of a magnetic field of 50 kOe suppresses the feature around 20 K, but a broad feature appears over a wider temperature range of 20-50 K (see figure 2d). Clearly, the imaginary part of *ac* susceptibility indicates a complex variation of spin-relaxation with the application of magnetic field. In short, if we compare *χ′(T)* for a particular frequency under different *dc* magnetic fields (as shown in figure 2e) for a fixed frequency of 133 Hz, it is clear that there is a shift of the peak position towards high temperature with increasing *dc* magnetic field, which is not very commonly known for a ferro/ferri magnetic system.

The ac susceptibility for other orientation of the crystal, *H⊥c*, does not show any feature (not shown here) which confirms the highly anisotropic spin glass nature of this compound.

**Complex dielectric behavior**

Now we will discuss the complex dielectric behavior for different crystallographic orientations of the single crystal with respect to electric field (*E*) and *H*.

Figure 3 shows the dielectric behavior for the orientation of *E//c//H* (the electric and magnetic fields are applied along the crystallographic '*c*' axis). There is a drop in the dielectric constant (ε′) as well as loss factor (tanδ) with decreasing temperature at the onset of long-range order, confirming magnetoelectric coupling. There is a ν-dependent behavior around 17-27 K for the frequency range 1-100 kHz (figures 3 a and e), as discussed for polycrystalline form of this compound (see Ref. 14 and 16); the magnitudes of ε′ are about an order of magnitude higher for this orientation compared to that for the polycrystalline form. However, the values of tanδ are rather small below 150 K, confirming insulating nature of the crystal for this orientation in the temperature range of investigation. The lowest frequency (1 kHz) in dielectric data is noisy but still one can infer that the peak temperature for this frequency in dielectric is comparable to that of highest frequency (1.3 kHz) used in ac susceptibility. Clearly strong ν-dependence suggests glassy electric dipole behavior of this compound as discussed for the polycrystalline compound earlier. Thus, the single crystal data confirms multiglass behavior of this compound, viewed together with *ac* χ behavior.

Now we discuss the effect of external magnetic field on the dielectric glassy behavior. It is clear from figures 3b-d that there is no change in the dielectric glassy behavior, even for an application of a high magnetic field of 50 kOe, in sharp contrast to the spin-glass response to *H* despite the presence of MDE coupling. This difference therefore suggests interesting dynamics of spins and dipoles, worth focusing on in future studies.

A point of emphasis is that, looking at the curves in the paramagnetic state, there is a broad feature around 70-130 K in $\varepsilon'(T)$ for *H*=0 (see figure 3a) at higher frequencies (see, for instance, for 100 kHz) similar to that seen in polycrystals of this compound [see Ref. 14]. This feature is somehow smeared at lower frequencies. Tanδ (figure 3e) also shows the broad-peak feature around this *T*-region for higher frequencies. This brad anomaly was attributed to incipient spin-chain ordering (14). Further increase in tanδ above 130 K could be due to the decreasing insulating nature with increasing temperature.

We have also investigated the dielectric behavior for the orientation of *E*⊥*c*//*H* (as shown in figure 4), where voltage is applied perpendicular to '*c*' and magnetic field is applied along the '*c*'. The observed frequency dependent dielectric behavior for the orientation of *E*⊥*c* (see figure 4a) is similar to that noted for the orientation *E*//*c*, though the feature is slightly weak and the values of dielectric constant are low compared to that for the orientation *E*//*c*. Interestingly, the dielectric glassy behavior is present for both the directions with respect to *E* (*E*//*c* and *E*⊥*c*), although the value of dielectric constant is different for these two directions. In short, if one looks at the absolute value of dielectric constant, the dielectric behavior is anisotropic. Returning to the dielectric glass behavior under external magnetic fields for the orientation of *E*⊥*c*//*H*, there is no change of dielectric feature for an application of magnetic field for this orientation as well (see figure 4 b-d for *H*= 0, 10 and 50 kOe). The low value of tanδ for this orientation as well supports highly insulating nature below around 150 K. Therefore, the dielectric glassy behavior observed at low temperatures is quite robust to the application of external magnetic field, despite the presence of strong magnetodielectric coupling (a behavior further discussed in the next section below). It is to be noted that quite similar robustness of dielectric behavior is observed in the isostructural compound $Ca_3Co_{1.6}Rh_{0.6}O_6$ in its single crystalline form [see Ref. 17]. Finally, the broad feature, observed around 60-130 K for the orientation *E//c*, is absent (or negligible) for this orientation (*E⊥c*). Presumably, since the magnetic chains are running along *c*-axis, such a feature is not observed for this orientation.

*Anisotropic MDE behavior*

Figure 5 shows the *H*-dependent dielectric behavior at different temperatures measured along the chain (*E//c//H*), in the form of $\Delta\varepsilon' = [\varepsilon'(H) - \varepsilon'(0)]/\varepsilon'(0)$ versus *H*. Below magnetic ordering, $\varepsilon'(H)$ clearly traces the magnetic field induced metamagnetic transition around 36 kOe (see figure 5 a-d), appearing in the form of a clear step around 36 kOe in $\varepsilon'(H)$, corresponding to the step in *M(H)* for *H//c* direction (see figure 1-c). This finding clearly confirms strong MDE coupling in this compound along the chain. It should be noted that the step in the plot of $\varepsilon'(H)$ in the range 10-24 K is at one-third of the highest-field (saturation) value, tracing the 'one-third' step in *M(H),* which suggests a 'one-to-one correlation' between magnetic and dielectric behavior of this compound. This finding (one-third step even in dielectric data) is fascinating. The magnetic hysteresis is also clearly observed due to MDE coupling (see figure 5a for 2.6 K). It is interesting that MDE coupling is present even far above magnetic ordering (see figure 5 e-f for 60 and 100 K), which clearly indicates that even short-range magnetic correlation can yield MDE coupling, as claimed for polycrystalline form [8, 9, 14]. Therefore, our results on single crystal confirm that one-dimensional short-range magnetic correlations can cause MDE coupling. It is clear from figure 5 c-f that the strength of MDE coupling decreases with increasing temperature. We have not studied MDE behavior at further higher *T* (> 130 K), as it is not of interest to us at this moment due to the less insulating nature of this compound.

Now, we will discuss MDE behavior for the orientation of *E⊥c//H* (shown in figure 6). In this orientation also, $\varepsilon'(H)$ clearly captures the magnetic field induced metamagnetic transition around 36 kOe at low *T*. However, the MDE coupling is negative in the forward cycle and also the values are nearly four times lower compared to that for the orientation *E//c//H*. Above $T_N$, say, for 62 and 100 K, MDE is weak, but with negative sign. Therefore,

MDE coupling is strongly anisotropic and the sign can be different for different orientations, namely, *E//c* and *E⊥c*.

We have also investigated the compound for other two orientations (*E//c⊥H* and *E⊥c⊥H*), but we did not find any appreciable MDE coupling (within the resolution limit of our instrument). This is also consistent with anisotropic MDE coupling.

*Magnetodielectric phase-co-existence*

Finally, we show the features attributable to novel phenomenon "magnetodielectric phase-co-existence" (a partial arrest of the dielectric phase at high magnetic fields). As described in polycrystalline form of this compound in our earlier publication [14], for single crystal also, we have observed similar dielectric arrest for the orientation of *E//c//H*; that is, after traversing the magnetic field across metamagnetic transition, the zero-field-state does not reach to its initial value even after sweeping the magnetic field back and forth for few cycles (see figure 5 a and c, for the curves at 2 and 5 K). If we cool the specimen under the application of a magnetic field (*H* > 36 kOe) from paramagnetic region (see figure 5b), this phenomenon is not observed as the initial state is already an arrested state, similar to that seen for polycrystalline form as well (Ref. 14). Such a phase co-existence is absent above $T_2$, say, at 20 K (see figure 5d). If we apply the electric field in the perpendicular direction to '*c*', but sweeping the magnetic field along the '*c*' (*E⊥c//H*), this phenomenon is observed for this geometry also (see the curves for 2.6 and 5 K in figure 6), but not above $T_2$. So, it is clear that "magnetodielectric phase-co-existence" is present along the spin-chain irrespective of the direction of applied electric field (for both *E//c* and *E⊥c*). This behavior is surprising as there is no such 'arrest' in isothermal magnetization *M(H)* (see figure 1c). The results therefore seems to reveal complex dynamics of spins and dipoles, despite the presence of strong MDE coupling, and its response to different probes (e.g., different time scales of relaxation and its hierarchy).

4. **Conclusion**

The results of magnetic, dielectric and magnetodielectric studies on the 'exotic' geometrically frustrated spin-chain compound $Ca_3Co_2O_6$ are presented for the single crystalline form in different orientations with respect to both electric and magnetic fields. Dielectric permittivity data tracks all the magnetic features for the orientation of *H//c*, establishing the existence of magnetodielectric coupling along the chain. The step in ε′(H) at one-third of highest-field (saturation) value for *H//c*, tracking the 'one-third' step in *M(H)*, suggests a 'one-to-one correlation' in magnetic and dielectric behavior of this compound. This should evoke theoretical interest. There is a huge frequency dependence of dielectric constant as a function of temperature as in the case of *ac* susceptibility below magnetic ordering. This glassy behavior of dielectric is robust to the applications of high magnetic-fields, unlike the situation in *ac* susceptibility. This suggests that the dynamics of electric dipoles and magnetic dipoles could be controlled differently by external parameters. The fascinating 'magnetodielectric phase-coexistence' phenomenon could be observed due to partial arrest of the high-field magnetodielectric phase, resulting in a butterfly-shaped loops in the plots of *ε′(H)* for *H//c* irrespective of the direction of applied *E*. Finally, there is a broad, but weak, peak in the range 60-130 K in the plot of ε′(T) for *H//c*, attributable to short-range correlation from 'spin-chain'.

**Aknowledgement:**

We would like to thank Dr. Kaustav Mukherjee for his participation during initial stages of single crystal preparation. We thank Dr. Kiran Singh for his help in dielectric measurements.

**Figure captions:**

Figure 1: Dc magnetization as a function of temperature for an application of 5 kOe magnetic field for the orientation of (a) *H//c* and (b) *H⊥c* of single crystal $Ca_3Co_2O_6$. Isothermal magnetization as a function of *H* at 2 K for (c) *H//c* and (d) *H⊥c* at 2 K.

Figure 2: Real and imaginary parts of ac susceptibility as a function of temperature for different frequencies (1.3 Hz-1.3 kHz) along $H//c$ of single crystal $Ca_3Co_2O_6$ in (a) the absence of *dc* magnetic field and in the presence of *dc* magnetic field of (b) 10 kOe, (c) 30 kOe and (d) 50 kOe. The temperature range is different in each plot to highlight respective features. The curves for two frequencies are deleted in (d) due to the noisy nature of the data and to clearly see the feature. (e) Real part of ac susceptibility as a function of temperature from 2-100 K at a fixed frequency of 133 Hz under different *dc* magnetic fields.

Figure 3: Dielectric constant as a function of temperature in (a) the absence of magnetic field and in the presence of magnetic field of (b) 10 kOe and (c) 50 kOe temperature for the orientation of $E//c//H$ of single crystal $Ca_3Co_2O_6$ for different frequencies (1, 10, 20, 30, 50, 70 and 100 kHz); (d) a comparison of the curves of $\varepsilon'(T)$ obtained in the presence of different magnetic fields; (e) loss factor( $\tan\delta$) as a function of $T$ for $H = 0$; 1kHz data is noisy and not shown here.

Figure 4: Dielectric constant as a function of temperature in (a) the absence of magnetic field and (b-c) in the presence of magnetic field of 10 kOe and 50 kOe, for the orientation of $E\perp c//H$ of single crystalline $Ca_3Co2O_6$ for different frequencies (1, 10, 30, 50 and 100 kHz); (d) a comparisons of the curves of $\varepsilon'(T)$ obtained in the presence of different magnetic fields; (e) loss factor( $\tan\delta$) as a function of $T$ for $H = 0$; 1kHz data is noisy and not shown here.

Figure 5: Magnetic field dependent isothermal dielectric behavior (change in dielectric constant as a function of $H$) at different temperatures (a) 2.6 K, (c) 5 K, (d) 20K, (e) 60 K and (f) 100 K, for zero-field-cooled condition; (b) 2.6 K for field-cooled condition with 140 kOe from 200 K, for a fixed frequency of 50 kHz, for the orientation of $E//c//H$ of single crystal $Ca_3Co_2O_6$. The numerical and arrows serve as guides to the eyes to show the sequence in which the magnetic field was changed.

Figure 6: Magnetic field dependent isothermal dielectric behavior (change in dielectric constant as a function of $H$) at different temperatures (a) 2.6 K, (b) 5 K, (c) 12 K, (d) 20K, (e) 60 K and (f) 100 K, for zero-field-cooled condition; (b) 2.6 K for field-cooled condition of 140 kOe from 200 K, for a fixed frequency of 50 kHz for the orientation of $E\perp c//H$ of single

crystal $Ca_3Co_2O_6$. The numerical and arrows serve as guides to the eyes to show the sequence in which the magnetic field was changed.

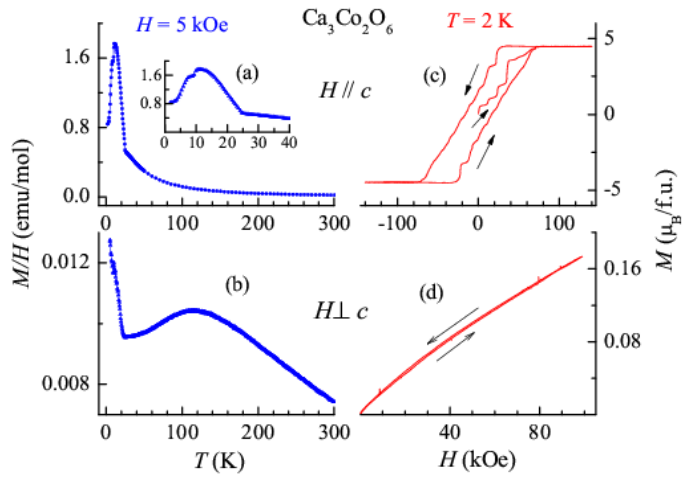

Figure 1.

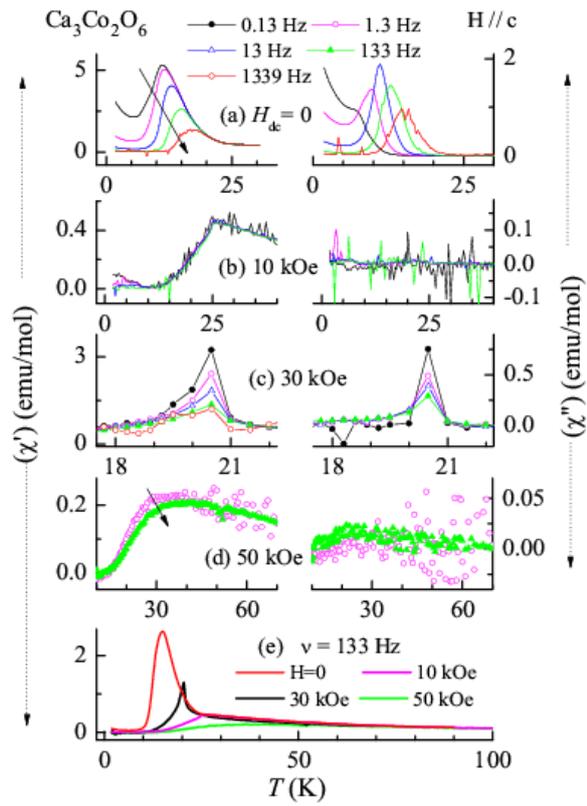

Figure 2.

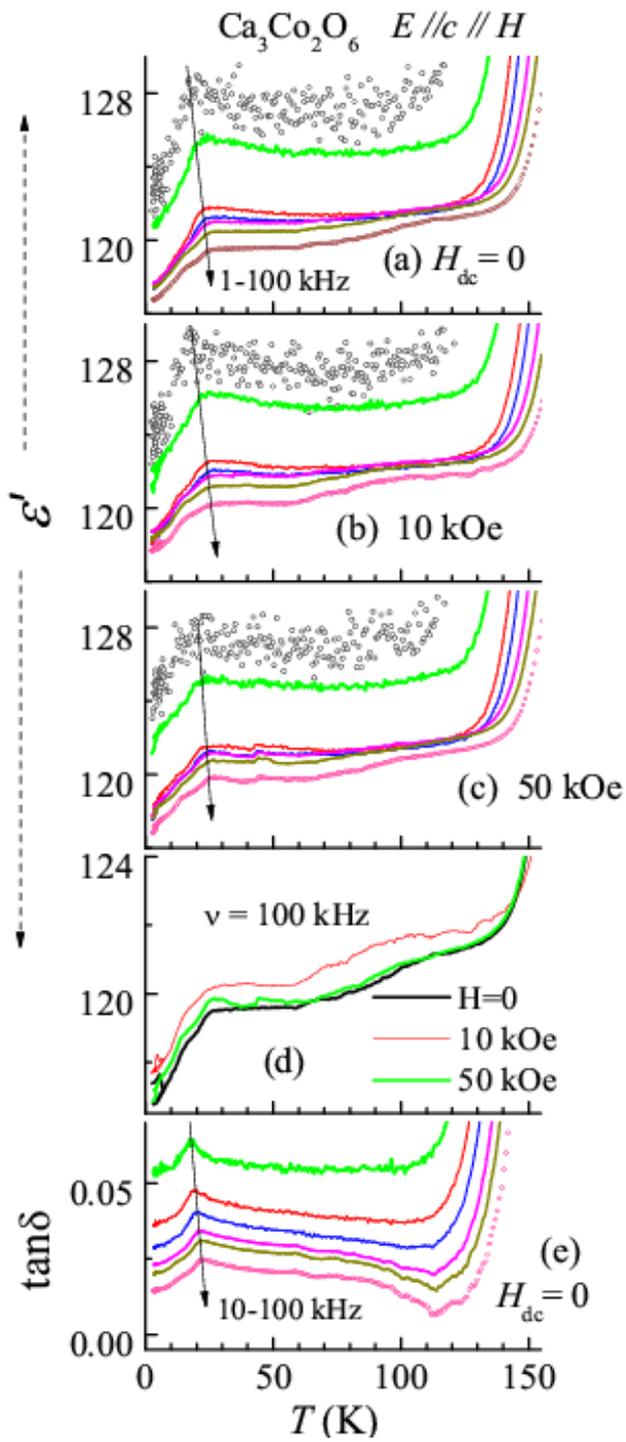

Figure 3.

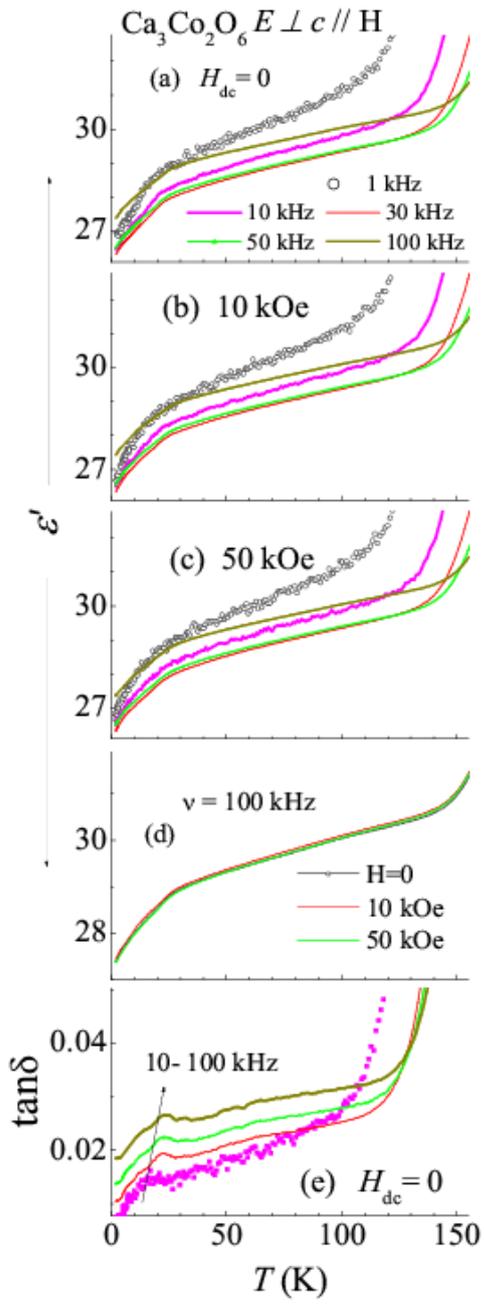

Figure 4.

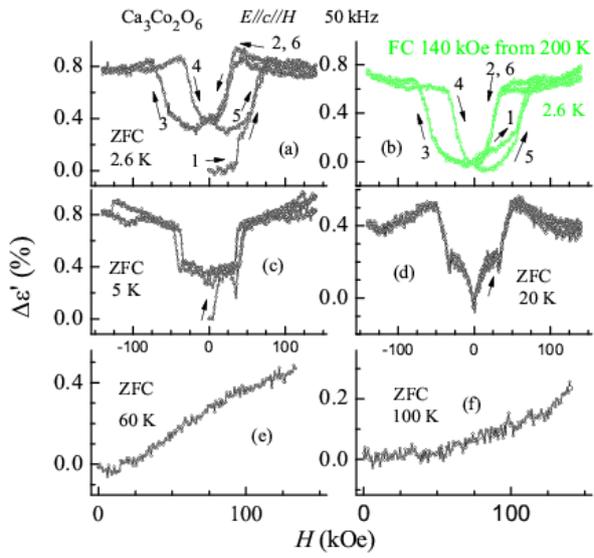

Figure 5.

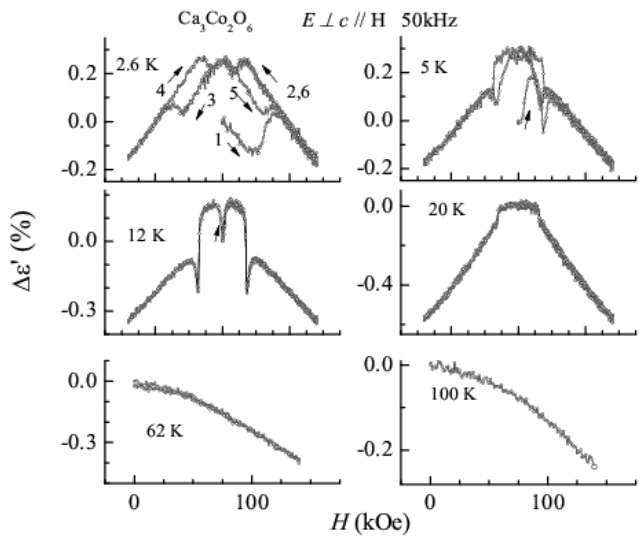

Figure 6.